\documentclass[a4paper]{article}

\usepackage[english]{babel}
\usepackage[utf8]{inputenc}
\usepackage{amsmath}
\usepackage{csquotes}
\usepackage{graphicx}
\usepackage[style=authoryear-ibid,backend=biber]{biblatex}

\providecommand{\keywords}[1]
{
  \small	
  \textbf{\textit{Keywords---}} #1
}

\title{How will AI and automation transform society and cities?}
\author{Gevorg Yeghikyan}
\date{\textit{Scuola Normale Superiore di Pisa} \\
\textit{ISTI-CNR} \\
gevorg.yeghikyan@sns.it }
\begin{document}

\maketitle
\begin{abstract}
    Against the backdrop of rising anxiety and discussions on the impact of AI on society, I explore in this article the structural possibilities of AI and automation triggering a new social conflict between the current capitalist elites and the emerging "creative class" (R\&D scientists, engineers, business developers, etc.), and how this conflict can produce social tensions and transform urban space. By drawing insights from a structurally similar conflict in 17-18th century Europe between the aristocracy and the emerging bourgeoisie, the impact of this conflict on the social, spatial, and power landscapes in cities of that time, as well as current trends in urban geography, this article outlines the prospects of urban transformations under changing production and consumption economies.  
\end{abstract}


\section{The prospects for a new social conflict}

In his classic work, the Production of Space, Henri Lefebvre argued that “new social relationships call for a new space, and vice versa.” \parencite{lefebvre1991production} With all the anxiety and discussions on AI and automation gaining momentum, the future of production relations, power structures, and social relationships becomes increasingly uncertain. Hand in hand, as urban theory suggests, so does the future of cities. To reflect on the possible urban futures unleashed by the effects of AI and automation on the social organization of space, one should ask, following Lefebvre: is a new social conflict structurally possible? The answer to this question might contain the seeds for ideas about the future of work, social relations, forms of social organization, and, ultimately, cities. 

\subsection{\textit{Formal} vs. \textit{real} subsumption}

So how should we approach this question? One need not be familiar with Marxist thought to know that capitalist production functions by appropriating the surplus value created by (social) labour. However, what is interesting regarding the possibility of a new social conflict due to AI and automation, is how capitalism subsumes labor. In Capital, Marx distinguishes between \textit{formal} and \textit{real} subsumption of labour by capital \parencite{marx2004capital}. Under \textit{formal} subsumption, capital exploits labour in socio-economic terms: the capitalist monopolises the means of production, forcing the worker to yield to the conditions of wage labour, be hired, and give away their added value. \textit{Real} subsumption, on the other hand, occurs along with the technological evolution of capitalist production, under which the worker is dominated by the capitalist not only by being hired, but also through exploitation by the material movement of capital in the process of production. He is dominated by the system of machines and becomes a kind of a living appendix to them, forced to work according to the technological logic of the machinery he operates, or rather, is operated by. In other words, the worker is subsumed by capital not only in terms of production relations, but also in terms of the very production processes he works under. This, according to Marx, is what makes the domination of labour by capital in organisational and administrative terms, and hence, the efficient obtaining of maximum surplus value from the employment of labour, possible. This theoretical framework can be easily extended to post-industrial economies dominated by services by replacing the machinery logic with, for instance, the bureaucratic-administrative logic of typical office work.

What is it that changes in capitalist production with the advent of Industry 4.0 with Machine to Machine, Internet of Things, and AI systems? The worker as an appendix to the production process essentially disappears, and with it, disappears also the bundle of manager and supervisor positions telling the worker what to do and how, since there is no one left to command: algorithms and robots are now at work. Unlike the disappearing factory or white collar workers, the remaining employees, required to develop, innovate, and sustain production, such as R\&D scientists, engineers, and business developers, are not working under \textit{real} subsumption by capital. They are mainly concerned with intellectual and creative work in which they are operationally free. Although the external socio-economic conditions are still dictated by capital since these workers are employed by it (\textit{formal} subsumption), they are not functionally dominated by capital in the creative process. Moreover, with the capitalist losing leverage over traditional workers due to automation replacing them, the creative workers gain further space for operational freedom. Thus, the widespread introduction of Industry 4.0 and AI systems completely or partially abolishes \textit{real} subsumption in capitalist production.

Interestingly, this is precisely where space for a new social conflict opens up. While the conflict between the traditional industrial worker and the capitalist revolved around wages, working hours and conditions, the creative worker is typically well-paid, enjoys retirement benefits, vacations, and is often even lured by employee stock options, offering company shares to its employees. Therefore, the conflict is to be found elsewhere.

\section{Social climbing in the 17th century and today}

To understand the possibilities of the creative class entering into conflict with existing capitalist structures, it is worthwhile to recall a structurally similar socio-economic conflict in history: that between the bourgeoisie and feudal aristocracy in 17th century Western Europe. Comedies of that epoch show what the emerging bourgeois was striving for: to make his way to the ruling classes through social climbing. His ascent from bourgeois to noble might involve marrying a noble woman or buying a noble title. For example, Moliere’s classic play, titled Le bourgeois gentilhomme, an evident oxymoron, mocks such attempts of social climbing by the pretentious bourgeois and ingeniously shows the emerging class conflict \parencite{moliere2019bourgeois}.

Likewise, being only \textit{formally} subsumed by capital and having considerable operational freedom, resources, and time at their disposal, \textbf{many creative workers today dream about converting their creative abilities into capital and applying them with profit - think of the startup ecosystems, - thus liberating themselves from the exploitation of capital and becoming capitalist entrepreneurs themselves}. As we can see, this is structurally analogous to the 17-18th century bourgeoisie striving to liberate themselves from the domination of the monarchic system offering privileges to the aristocracy, by attempting social climbing to become aristocrats themselves. This state of affairs continued for almost two centuries, until the conflict between the ever-stronger bourgeoisie and the ever-weaker aristocracy first manifested itself in the former articulating their differences from the latter, and then burst into the violent bourgeois revolution.

\subsection{Limits to the growth of the creative class}

Coming back to our days, capital still succeeds in keeping the ever-growing creative class \parencite{Florida_creative} in check by offering privileged working conditions and employee stock options to a still narrow group of creative workers, but it will not be able to do so indefinitely. Take the example of the so-called Scandinavian Nordic model. The Nordic model, more than any other capitalist model, took the path of collective bargaining: a strong compromise between capital and labour, with an unparalleled level of social guarantees, education and healthcare in the developed part of the world. For example, unlike the United States where one in ten people live officially below the poverty line, poverty in the Scandinavian countries is virtually nonexistent \parencite{poverty_US}. Similarly, a situation where a person is denied medical care because they cannot afford it is essentially impossible in those countries \parencite{denied_medicare}. Nevertheless, the Nordic model has its limits and reached them already in the 1980s when the portion of GDP redistributed through government spending, i.e., through a non-market mechanism, reached 65\% \parencite{owidgovernmentspending}. This meant that capital was forced to concede the lion’s share of the surplus value it appropriated from labour and allow it to be redistributed among workers. The Nordic model was unable to go further with its redistributive policies, since doing so would mean to encroach the very heart of capitalist production denying it its main motivation: the appropriation of surplus value. Similarly, given the current growth of the creative class, capital will not be able to attract and sustain \textit{all} creative workers, and, given the discussed aspirations of the latter, the conflict will thus be only exacerbated.

Now, whether the discussed conflict between the growing creative class and the current bourgeoisie will culminate in a revolution or will find another resolution is not only difficult to predict, but is secondary and irrelevant here. Of interest is how it can affect the social, political, and economic organization of urban space.

\section{Urban transformations in 19th century France}

To understand the prospects of these pending socio-economic transformations shaping the urban landscape, let’s have a recourse to the French Revolution, and its short and long term aftermath. Before and immediately after the revolution, in the late 18th century, the aristocracy, bourgeoisie, artisans, and workers had different levels of urban mobility and use of urban space. Municipal power, exercised by the monarchy and the aristocracy tightly connected with it, built elaborate public spaces for displaying power as a performative ritual \parencite{rotenberg2001metropolitanism}: processions of carriages pompously moved every day along stylised boulevards connecting the various “h\^{o}tel particulier” to municipal or religious buildings situated in focal squares. Place des Vosges and Place de la Concorde in Paris are prime examples of such urban spaces. Thus, \textbf{public space didn’t exist in the modern sense and merely served as a stage for celebrating monarchic power}. While lower class workers and artisans exercised minimal mobility staying within the confines of their neighbourhoods and travelling almost exclusively by foot, the emerging bourgeoisie was increasingly mobile, agile, and sought to establish its class identity. Class membership was first articulated through literacy, and a new consumer culture such as the emergence of restaurants as a cheaper way to simulate a private cook that only nobles could enjoy \parencite{spang2019invention}.

And yet, segregation in cities was mainly vertical rather than horizontal, meaning that the poor lived on the higher floors of buildings inhabited by the upper classes \parencite{chartier1998ville}. One could often meet middle class members, artisanal masters, and lower class workers in the same neighbourhood. This was due to the initial alliance of reformist middle classes with the relatively well-doing artisanal classes.

This began to change with the July revolution of 1830, when space started to play a central role in the creation of middle-class identity following similar processes in London and Vienna. The noble landowners had overestimated the falling demand for noble mansion housing, incurring severe losses on investment, and eventually had to sell the land to private investors. The increasing migration to cities and the ensuing severe housing crisis, coupled with the socio-economic transformations with a growing bourgeoisie and its quest for a new identity, compelled private developers to create and sell an image of a culturally homogeneous, technologically modern, and consumer-quality neighbourhoods in order to make money on new housing. Private capital thus subdivided previously aristocratic lands into smaller lots and made possible urban redevelopment in specific areas of the city, meeting the newly formed bourgeois consumer demand. These new quarters instigated the self-reinforcing horizontal stratification of the city, with rising property values attracting financial and banking elites and the upper middle class, which further nourished consumer culture, already manifest in the design of new housing blocks.

To distinguish themselves both from the aristocracy and the lower classes, new kinds of public and private spaces were emerging. This included the appearance of boudoirs and, most notably, balconies, which, as argued by Francois Loyer, were aimed at displaying public wealth by “showing the world outside the location and size of an apartment’s salon" \parencite{loyer1988paris}. The streets in these new quarters were wider, straighter, and more luminous, containing more parks and public spaces. The socio-economic conditions thus shaped a different kind of urban space, which made it possible to experience older areas of the city as well. As Victoria Thompson argued, the experience of wider streets made older streets seem even narrower, and this came to be associated with the poor, dark neighbourhoods \parencite{thompson2003telling}.

This only strengthened segregation in the city, and, given the low urban mobility of the lower classes and the poor living conditions in such centrally located older parts of the city as Cité in Paris, planted the seeds for social tensions between the poor and the bourgeoisie for the first time. For example, as recounted by Eugene Roche, it was common belief among the poor in these quarters that the 1832 cholera was a plot of the rich (sounds familiar?), leading to disregard for social distancing and hygiene rules, and propelling insurrection, resulting in a mortality rate in the poor quarters twice as high as the city average \parencite{roch1832paris, jordan1995transforming}.

Bourgeois urban redevelopment, the emergence of what we now call urban master planning, the Hausmann reconstruction, and rising rents eventually, towards the end of the 19th century, pushed the poor away from the city center to the city outskirts. Here, the poor working class lived in badly built, unsanitary housing and environment, from where the poor did not intrude the rich bourgeois quarters anymore. This concluded the horizontal segregation process that had begun almost a century earlier.

\section{Customizing or segregating urban space}

This rather lengthy historical excursion served to delineate the framework for a conceptually similar possible transformation of urban space following the emergence of the creative class and its conflict with the current bourgeois elites. The economic geography for one such transformation model has been described by the urbanist Dror Poleg in a recent article \parencite{living_tail}.  Poleg starts by recounting the following idea of Chris Anderson: historically, the entertainment industry was built around producing and promoting "hits" because of economic and physical restrictions. The production, shipping, storing, and displaying CDs or books were expensive, there were not many cinemas, and there were only 24 hours in a day. This meant that producing a wide range of albums, books, or movies was not economically viable and hence the industry was dependent on hits following the Pareto rule: 20\% of products made 80\% of the sales \parencite{anderson2007long}.

\begin{figure}[ht]
    \centering
    \includegraphics[width=0.9\linewidth]{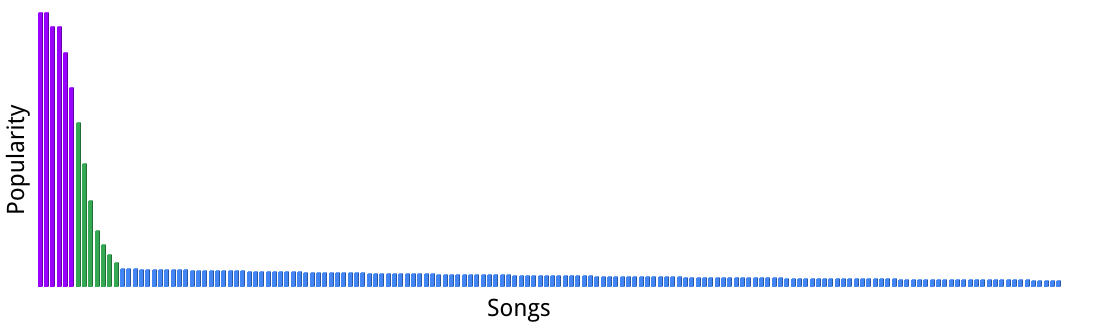}
    \caption{Song popularity vs. sales rank plot diagram. Copyright Dror Poleg.}
\label{Poleg}
\end{figure}

The internet changed this by making the production, distribution and storage of, and access to content cheaper. Now, most of the sales are no longer coming from a few hits, but from what Anderson calls the "long tail" of a multitude of highly customized products, of interest to a very narrow audience, as shown in Figure \ref{Poleg}.

The sheer quantity of such "narrow" products creates near-unlimited choice and constitutes the bulk of most consumer demand. Poleg calls this phenomenon the economics of abundance and goes on to link this idea to urban geography. \textbf{While economic scarcity historically produced spatially concentrated population and economic activity distributions in cities, the emergence of remote working options is causing population and economic activity to be distributed across a long tail of many city locations instead of a few spatially concentrated centers as before}. He then notes that under economics of scarcity, the spatially concentrated urban locations were designed to appeal to the largest number of people since they share the same urban infrastructure, centrally located job market, and public spaces. However, under economics of abundance, people won’t have to share the same space anymore. \textbf{City locations could become as customized as the songs people choose to listen to, making cities even more segregated}.

\section{So what is at stake?}

In the developed world, given the growing ambitions and aspirations of creative workers, gradual and unequal automation can trigger new forms of social organisation of space following the logic discussed in our historical excursion as well as in the shift of urban activity to the "long tail". In fact, the impact of the growing political relevance of the creative class on real estate is already palpable in such business models as that of WeWork. As described in a piece in the Atlantic \parencite{wework}: 

"\textit{The bespoke spaces are appealing to big organizations because they allow for a nimble office experience in which employees aren’t necessarily bound to their desks but are free to roam and connect with other businesses and entrepreneurs, creating a looser, more innovative environment in which ideas can sprout.}"

Going further, \textbf{urban segregation on the "long tail", for example, can take the form of a digital divide, but in reverse}. With automation bringing low skilled labour wages below the level of subsistence, the unemployed will become structurally unemployable, and universal basic income (UBI) will most likely become an imperative policy to sustain this segment of population, making it a class of its own. This \textbf{UBI} class will be offered basic services in their neighbourhoods and cheap access to digital services, but will be denied a range of new kinds of physical services like novel infrastructures serving specific technologies in transport, logistical, and recreational activities, as well as quality healthcare and education in customized quarters of the creative classes. \textbf{This might create a feedback loop of further investment in those areas, creating new extraterritorial conditions, pushing the UBI class away, thus concluding the urban segregation process, just like the ascending bourgeois pushed low wage workers away to the city outskirts 150 years ago}.

Moreover, a certain physical-digital consumer culture of minor luxuries will have to be facilitated to keep the UBI class entertained within the range of their basic income, further encouraging this class to customize their neighbourhoods according to that culture. The closest analogy is perhaps the story of the French general Gallieni who conquered Madagascar, imposed a head tax in the newly issued local currency in 1901, effectively printing money and demanding that everyone give some of that money back to him. Described in detail in David Graeber’s seminal work, this forced the creation of certain markets and population debt, but was also aimed at nurturing a consumer culture of cheap luxuries like lipstick or parasol, and making sure locals have some of their money left to spend on these products \parencite{graeber2012debt}. It was crucial that they develop new tastes, habits, and desires; that consumer demand previously alien to them became ingrained in local culture to keep Madagascar forever tied to France even after the conquerors had left.
Despite the speculative nature of these reflections, they highlight the possibility for social and urban transformations articulated through extraterritorial conditions and segregation in our cities, as a result of the emerging structural conflict between the creative class and the current capitalist bourgeoisie.

\subsection{Hot dog urbanism}

One key aspect of our discussion about automation and AI missing so far is that automation is not taking place uniformly across all levels of production. This consideration is based on Paul Krugman's hot dog economy toy model \parencite{krugman2010accidental}. In this model, the economy produces just hot dogs, consisting of two goods, sausages and bread rolls. Let’s assume that advances in technology allow the bakeries to be automated, boosting productivity and dropping the cost per bread roll produced. This cost reduction causes a rise in demand for hot-dogs, other things being equal. But since automation took place only in bakeries, while sausages continue to be produced manually, this means that demand for manual labour in the sausage industry will have to rise to meet the growing hot-dog demand.

So far, the solution has been to tap into cheap labour in the third world. For example, a technological innovation in Silicon Valley increases productivity in the upper levels of production but simultaneously raises demand for additional manual labour in the lower levels of production. Consider that despite automation and AI, the global workforce has actually doubled in the past decades reaching 3.5 billion people - faster than population growth \parencite{labour_force}. Incidentally, this might also be the reason why despite the technology being available for quite some time now, corporations have not been in a hurry to shift to automated production, preferring cheap manual labour somewhere in Southeast Asia instead. However, unless automation takes place uniformly across different levels of production chains, which it won’t, we might actually face a workforce shortage, given the limited rural population and the current rates of urbanization. This means that the above mentioned changes in social structures and urban life will have an unequal distribution on the global scale with inevitable geopolitical consequences, which is, however, out of the scope of this article.

\section{Conclusion}

In an attempt to sketch the baseline for how AI and automation could affect social and urban structures, we looked at concepts from Marx’s political economy, saw what kind of social conflict could emerge with the advent of automation, and looked at 17-19th century France to discover how the mounting socio-economic conflict between the aristocracy and bourgeoisie shaped the urban landscape of Paris and resulted in segregation in the city. We discussed Dror Poleg’s idea of living on the tail by noting how it structurally fits the predictions of our model of the social conflict between the creative class and current capitalist elites. We also speculated about the formation of a structurally unemployable class sustained by universal basic income, the \textbf{UBI class}, and reflected on possible urban transformations as an indispensable part of this dynamic.

City authorities have traditionally been \textit{reactive}, but with the advent of the Information age and the disruptive power of AI and automation, policy makers will be compelled to take a \textit{proactive} approach to addressing the challenges to the social and cultural fabric raised by the impact of new technologies on urban space.

This discussed scenario is, no doubt,  highly speculative, and the exact form it could take is impossible to predict, but the aim was to draw attention to the new structural conflict emerging in the current post-industrial capitalist regime and to trigger further discussions on how it might transform society and urban space.

\printbibliography

\end{document}